\documentclass[10pt,conference,a4paper]{IEEEtran}

\renewcommand\IEEEkeywordsname{Keywords}

\usepackage{amsmath} 
\usepackage{amsfonts} 
\usepackage{amssymb}
\usepackage{amsthm}
\usepackage[utf8]{inputenc}
\usepackage{subfigure}
\usepackage{caption}
\usepackage{bbm}
\usepackage{csquotes}

\ifCLASSINFOpdf 
\usepackage[pdftex]{graphicx} 
\usepackage[pdftex,colorlinks,
           bookmarks=true,%
            pdftitle=Wireless\ Node\ Cooperation\ with\ Resource\ Availability\ Constraints,%
           pdfauthor= \ L.D.\ Alvarez Corrales -A.\ Giovanidis\ -\ P.\ Martins -\ L.\ Decreusefond]{hyperref}  
\usepackage[numbers,sort&compress]{natbib} 

\DeclareMathOperator*{\argmin}{argmin}

\newtheorem{defi}{Definition}
\newtheorem{prop}{Proposition}

\newtheorem{theo}{Theorem}
\newtheorem{lem}{Lemma}

\begin{document}

\title{Wireless Node Cooperation with Resource Availability Constraints}

\author{Luis \'Alvarez-Corrales$^1$, Anastasios Giovanidis$^2$, Philippe Martins$^1$, and Laurent Decreusefond$^1$\\
$^1$ LTCI, T\'el\'ecom ParisTech, Universit\'e Paris-Saclay, Paris, France \\
$^2$ CNRS-LIP6, Universit\'e Pierre et Marie Curie, Sorbonne Universit\'es, Paris, France \\
Contact: $^1\{$firstname.lastname@telecom-paristech.fr$\}$, $^2\{$firstname.lastname$\}$@lip6.fr
}

\maketitle

\begin{abstract}
Base station cooperation is a promising scheme to improve network performance for next generation cellular networks. Up to this point research has focused on station grouping criteria based solely on geographic proximity. However, for the cooperation to be meaningful, each station participating in a group should have sufficient available resources to share with others. In this work we consider an alternative grouping criterion based on a distance that considers both geographic proximity and available resources of the stations. When the network is modelled by a Poisson Point Process, we derive analytical formulas on the proportion of cooperative pairs or single stations, and the expected sum interference from each of the groups. The results illustrate that cooperation gains strongly depend on the distribution of available resources over the network.
\end{abstract}


\section{Introduction}
Cooperation between base stations (BSs) is a topic of considerable ongoing research in cellular networks. It is particularly beneficial for users located at the cell-edge. The cooperation concept is expected to play a significant role in future planning and deployment, due to the coming densification of networks by HetNets \cite{DhillonBest12}. There is a considerable amount of research on the topic, related to the concept of CoMP \cite{TheRolSmallCellsJungV}, Network MIMO \cite{GesMultMIMO2010,GioA012012}, or C-RAN \cite{CloudRANChecko2015}. The different cooperation methodologies proposed differ in the way groups of BSs are formed, the number of cooperating nodes, the type of signal cooperation, and the amount of information exchange. In \cite{BacAStoGeo2015,NigCoordMul2014,BlasStuSINTFact2015,TanTracMod2014}, the authors introduce clustering methodologies where the user \textit{dynamically selects} the set of stations cooperating for its service.  Other authors propose to group BSs \textit{in a static way}, independently of the user configuration \cite{AkouIntCoor2013,ParkColBSs,AStocGemMultCellHuang2011,GuoSPGP2014,AfshFundClustCent}. 

Modeling wireless networks via stochastic geometry allows to consider the irregularity of the BS locations, as well as the randomness of other parameters of the telecommunication network (e.g. fading, shadowing), on the users' performance \cite{AndATract2011,BacBlaVol1}. Using this tool, the gains from BS-cooperation can be quantified systematically, so there is no need to test each different instance of the network by simulations. Closed formulas are very important for an operator that wants to plan and deploy infrastructure with cooperation functionality \cite{BacAStoGeo2015,BlasStuSINTFact2015,GuoSPGP2014}.

\subsection{The Mutually Nearest Neighbor Relation}

\begin{figure}[htbp] 
\centering
\subfigure[]{\includegraphics[trim = 10mm 4mm 10mm 2mm, clip, width=170pt]{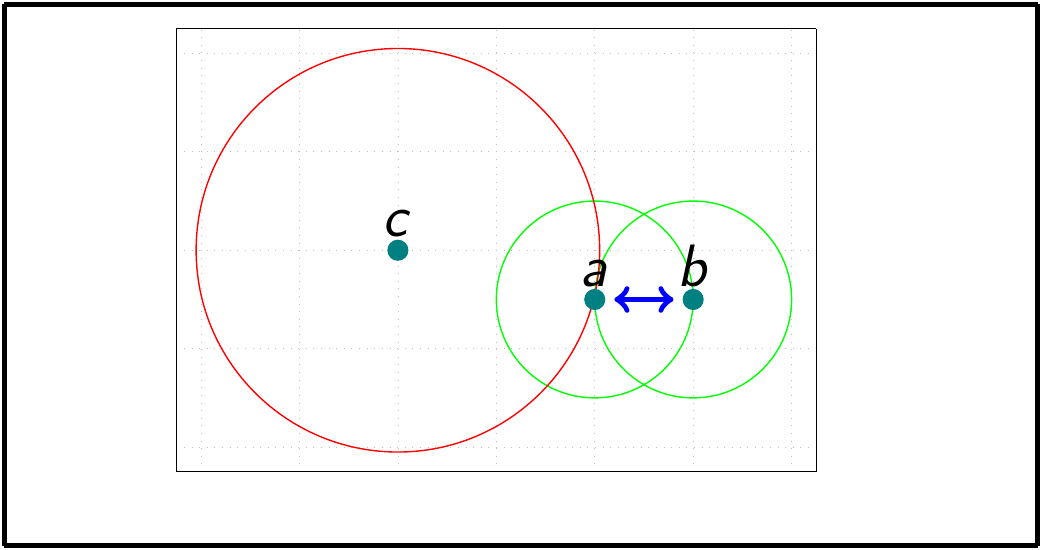}\label{MNNRa}}
\subfigure[]{\includegraphics[trim = 10mm 4mm 10mm 2mm, clip, width=170pt]{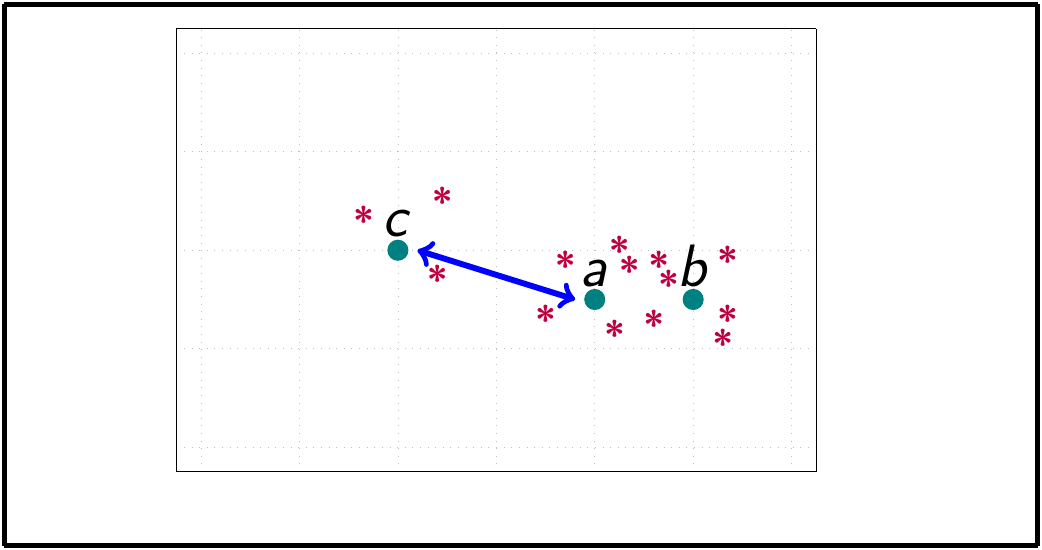}\label{MNNRb}} 
\caption{(a) The atom $a$ is the nearest neighbor of $b$, and $b$ is the nearest neighbor of $a$. Hence, we say that they are in MNNR, and they can work in pair. The atom $a$ is the nearest neighbor of $c$, but $c$ is not the nearest neighbor of $a$. Thus, $c$ is a single. (b) If $a$ and $b$ serve a large number of users (asterisks), the available remaining resources of both are low. If $c$ does not serve as many users, it has a considerable amount of unused resources. Since $c$ is close to $a$, their alternative cooperation can allow users of $a$ to be partly served from both $a$ and $c$, while the signals from $c$ can be sufficiently strong, due to proximity. This example ilustrates that the Euclidean proximity between the BS locations is not enough to ensure efficient cooperation.} 
\label{MNNR}
\end{figure}

In \cite{GioAnalInt2015,CovGainsAlv2016} the authors proposed a \textit{static clustering criterion}, where two atoms belong to the same cooperating group if each one of them is the geographically nearest neighbor of the other. We say that they are in \textit{Mutually Nearest Neighbor Relation (MNNR)}. On the other hand, if the nearest neighbor of one atom has a different nearest neighbor, then, the former atom remains single (see Figure \ref{MNNRa}). 
Given a fixed deployment of BSs, this criterion either allows BSs to form cooperative pairs, or to remain single operating on their own. This grouping method, which is based on Euclidean proximity, succeeds to reduce interference at the user side, with minimum additional infrastructure (only communication links between the cooperating BSs are required to be installed). However, the criterion does not take into account whether the members of each cooperative pair have sufficient resources, in order for their cooperation to be beneficial for the users (see a motivating example in Figure \ref{MNNRb}). 

\subsection{The MNNR with resource constrains}
Each BS has a certain resource availability, which can be quantified by a a positive mark. This mark can be the amount of available resources itself (unused bandwidth, unoccupied OFDMA slots, etc), or a system indicator such as the residual capacity (see \cite{RobMethLau2012}, and Figure \ref{MNNRb}) or the coverage (one-minus-outage) probability \cite{DhillonBest12,BacAStoGeo2015,NigCoordMul2014,TanTracMod2014,BlasStuSINTFact2015,AndATract2011}. The higher the mark, the more available the BS is to serve additional users. This resource availability mark will be made specific in each example and simulations environment that follow.  

Our aim in this work is to propose an extension of the MNNR, which allows the formation of clusters between BSs, that have the following two properties:
\begin{itemize}
\item[(1)] their locations are \textit{Euclideanly} close, and,
\item[(2)] they have sufficient resources for the cooperation to be beneficial.
\end{itemize}  
In the mathematical model, a BS is characterised by a 2-dimensional location, along with a positive mark (its available resources). Thus, we consider each BS in the 3-dimensional Euclidean space. As a result, we need to adjust the MNNR criterion in this space, by using a distance that takes into consideration both, location and available resources, appropriately. 

The natural choice would be the 3-dimensional Euclidean distance. This distance, however, would allow cooperative pairs to be formed (whenever their location are geographically close, in the 2-dimensional sense), whose BSs can both have \textit{an arbitrarily small amount} of available resources. Thus, the resulting cooperative pairs would not be reasonable for our engineering purposes. 

On the other hand, the 3-dimensional hyperbolic distance is actually a good candidate for our particular aim. Indeed, grouping the BSs according to their relative hyperbolic proximity (to be defined precisely in what follows) translates straightforwardly into network benefits: For the BSs in a cooperative pair, there is interference related improvements (as shown in \cite{GioAnalInt2015,CovGainsAlv2016}), while both BSs make use of their common available resources to serve the combined user load, implementing a type of \textit{load balancing}. Such cooperative pairs can be used within the framework of C-RAN. 

We will see that, when the BS locations are modeled by an independently marked Poisson Point Process (PPP), the hyperbolic distance makes the derivation of interesting analytic results possible. 

\textit{Hyperbolic Geometry} was developed in the 19th century as an alternative to the discussion about the parallel postulate of Euclidean Geometry. 
During the 20th century, physicists found in it convenient tools to work in the fields of \textit{mathematical physics} and \textit{special relativity} \cite{GlobLorenBeem1996}. Nowadays, other applied fields such as \textit{mathematical finance} and \textit{option pricing} benefits also from it \cite{AnalGeoHen2008}. In recent years, important research in \textit{communication networks} \cite{HypTrafStai2016}, \textit{complex networks} \cite{ComNetHyp}, and \textit{big network data} \cite{AHypBigDatStai2016} has also found in hyperbolic metric spaces an approach aspiring to radically change current practices. 

As a remark, the grouping criterion proposed in this work is neither entirely \textit{dynamic}, nor entirely \textit{static}, as the clusters can change to adjust to the resources of the BSs.

\subsection{Contributions}
This paper provides the following contributions:

\begin{itemize}
\item In Section \ref{SecII}, we introduce the hyperbolic distance between two \textit{marked} nodes, and further elaborate on its properties. The Mutually Nearest Neighbor Relation (MNNR) of \cite{GioAnalInt2015,CovGainsAlv2016} is redefined in the hyperbolic half-space, so that the resource availability is taken into account in the formation of cooperative groups. The hyperbolic MNNR criterion splits the BSs into cooperative pairs and singles. By means of example, we explain how an operator can adjust the MNNR criterion to control cooperation.
\item In Section \ref{SecIII}, we analyze hyperbolic MNNR cooperation when BSs are modeled by a marked PPP. Specifically, we derive an explicit expression for the probability of two given BSs to cooperate (Theorem \ref{theo1}). We further provide integral representations for the expected value of the interference generated by the singles and by the cooperative pairs (Theorem \ref{ExpSin}). These formulas hold for general signal transmission strategies of cooperation/coordination. Finally, an explicit expression for the average number of cooperative pairs in the network is provided (Theorem \ref{proporDoub}). 
\item In Section \ref{SecIV}, we show how the percentage of cooperative pairs depends on the distribution of resources. We compare this percentage with the one from the purely geometric model in \cite{GioAnalInt2015,CovGainsAlv2016}. Additionally, the numerical evaluation of the expected interference formulas is illustrated in plots and compared with simulations. 
\item Section \ref{SecV} draws the general conclusions of the analysis. 
\end{itemize}
Proofs of  Lemmas, Propositions, Theorems, and other facts can be found in the Appendix. 

\section{The Mutually Nearest Neighbor Relation in $\mathbb{H}^3$ and hyperbolic proximity}\label{SecII}
\subsection{Notation}

Let $\mathbb{H}^3=\{(x,y,z)\in \mathbb{R}^3 \ | \ z>0\}$ denote the three dimensional hyperbolic half-space. The letters $a$, $b$, and $c$ will denote atoms (BSs) in the hyperbolic space $\mathbb{H}^3$. 

For some atom $a=(x,y,z)$ in $\mathbb{H}^3$, the vector $(x,y)$ is the 2-dimensional Euclidean position of $a$, and $z>0$ are its resources. We use the projection map 
\begin{equation*}
\begin{split}
\hat{\cdot}:\mathbb{H}^3 & \rightarrow \mathbb{R}^2 \\
a &\mapsto \hat{a}=(x,y)
\end{split}
\end{equation*}
to obtain the 2-dimensional Euclidean position of every atom in $\mathbb{H}^3$. In this fashion, the letters $\hat{a}$, $\hat{b}$, and $\hat{c}$ will denote elements of $\mathbb{R}^2$.

For every two atoms $a=(x,y,z)$ and $b=(\tilde{x},\tilde{y},\tilde{z})$, denote $$d_E(a,b):=\sqrt{(x- \tilde{x})^2+(y- \tilde{y})^2},$$ the 2-dimensional Euclidean distance between $a$ and $b$.

Let $\mathcal{B}(\mathbb{R}^2)$ be the set of Borel-measurable subsets of $\mathbb{R}^2$. The Lebesgue measure of the subset $A\in \mathcal{B}(\mathbb{R}^2)$ is denoted by $\mathcal{S}(A)$.

\subsection{Hyperbolic geometry tools}\label{proximity}

Let $a$ and $b$ be two atoms  in $\mathbb{H}^3$, with resources $z$ and $\tilde{z}$, respectively. The hyperbolic distance between $a$ and $b$ is given by the expression
\begin{equation}
\label{hyperDist}
d_{\mathbb{H}^3}(a,b) := acosh\left(\frac{d_E(a,b)^2}{2z \tilde{z}} + \frac{1}{2}\left( \frac{z}{\tilde{z}}+\frac{\tilde{z}}{z}\right) \right),
\end{equation}
where $acosh(\cdot)$ denotes the inverse of the hyperbolic cosine function \cite[Prop.1.6]{GroActHypSpa1998}. 

\subsubsection{Hyperbolic proximity in terms of Euclidean proximity and resources}
As stated in the introduction, we will create the clusters with respect to hyperbolic proximity. First of all, let us analyse what it means for two atoms $a$ and $b$, whose resources are $z>0$ and $\tilde{z}>0$, to be hyperbolically close. For the two atoms to be close in $\mathbb{H}^3$, it is sufficient to analyse when the expression 
$\frac{d_E(a,b)^2}{2z \tilde{z}} + \frac{1}{2}\left( \frac{z}{\tilde{z}}+\frac{\tilde{z}}{z}\right)$
is small (because $acosh(\cdot)$ is an increasing function). 

The continuous function
\begin{equation*}
(z,\tilde{z},d_E(a,b)) \longmapsto \frac{d_E(a,b)^2}{2z \tilde{z}} 
\end{equation*}
attains its minimum value $0$ at $\{(z,\tilde{z},0) \ | \ z,\tilde{z}>0 \}$. Hence, given $z$ and $\tilde{z}$, this function is close to the minimum, when the values of $d_E(a,b)$ are small. Given $d_E(a,b)$, the function is close to the minimum as well, when the values of the product $z \tilde{z}$ are large (the resource indicator of one of the BSs or both are large). As a remark, notice that, for a given $d_E(a,b)$, the previous function explodes whenever the product $z \tilde{z}$ is small. In telecommunication terms, this means that cooperation between two nodes that do not have enough resources is not favorable (\textit{i.e.}, the resource indicator of one-out-of-two, or both, is small).

From the second term, the continuous function 
\begin{equation*}
(z,\tilde{z}) \longmapsto \frac{1}{2}\left( \frac{z}{\tilde{z}}+\frac{\tilde{z}}{z}\right)
\end{equation*} 
attains its global minimum value $1$ at $\{ (z,z) \ | \ z>0 \}$. Therefore, whenever $z \approx \tilde{z}$, the above function is close to its minimum. Thus, only the pairs of BSs whose resources are in balance would be candidates for cooperation. This term enforces a sort of \textit{load balancing} among the BSs in cooperation. 

With the above discussion, we summarize that two atoms $a$ and $b$ are hyperbolically close if all the three following conditions are satisfied: 
\begin{itemize}
\item[(1)] they are geographically close (in the 2-dimensional sense),
\item[(2)] the product of their resource indicators is not small,
\item[(3)] the quantities of available resources of both are balanced.
\end{itemize}

\subsubsection{A key property of the Hyperbolic half-space}

For $a\in \mathbb{H}^3$, and $\epsilon>0$, let  
\begin{equation*}
B_{\mathbb{H}^3}(\epsilon,a):=\{ b\in \mathbb{H}^3 \ | \ d_{\mathbb{H}^3}(a,b)<\epsilon \}
\end{equation*}  
be the hyperbolic ball, centered at $a$, with radius $\epsilon>0$.  

\begin{defi}\label{EucBall}
Suppose that $a=(x,y,z)$. We denote by $B_E(\epsilon,a)$ the 3-dimensional Euclidean ball, centered at $(x,y,zcosh(\epsilon))$, with radius $zsinh(\epsilon)$, where $sinh(\cdot)$ is the hyperbolic sine function.
\end{defi}
We have the following result.

\begin{prop}\label{prop1}
The hyperbolic ball $B_{\mathbb{H}^3}(\epsilon,a)$ is described in the 3-dimensional Euclidean space by $B_E(\epsilon,a)$.
\end{prop}

\subsection{The Mutually Nearest Neighbor Relation in $\mathbb{H}^3$}
Every configuration of points $\phi$ over $\mathbb{H}^3$ represents a given topology for the BS locations, and given resource indicator for each one of them. Let $\phi$ be a simple, locally-finite, configuration. For $a$ and $b$, two different atoms in $\phi$, we say that $a$ is in \textit{Nearest Neighbor Relation in $\mathbb{H}^3$ (NNR$_{\mathbb{H}^3}$)} with $b$, with respect to $\phi$, if 
\begin{equation*}
b := \argmin_{c \in \phi\backslash \{a\} }d_{\mathbb{H}^3}(a,c),
\end{equation*} 
and we write $a \stackrel{\phi}{\rightarrow} b$. When $a$ is not in NNR$_{\mathbb{H}^3}$ with $b$, we write $a \stackrel{\phi}{\nrightarrow} b$. 

Henceforth, we will only consider configurations fulfilling the uniqueness of the nearest neighbor. Even if this uniqueness is not true in general, when the atoms are modeled by a stationary point process, this condition holds $\mathbb{P}$-a.s. (recall that every stationary point process has the Lebesgue measure as its intensity measure, see the Appendix). 

Consider the set $\mathcal{D}\subset (0,\infty)\times (0,\infty)$, which is Borel measurable and symmetric (\textit{i.e.}, if $(z,\tilde{z})\in \mathcal{D}$ then $(\tilde{z},z)\in \mathcal{D}$). This set will allow to control the creation of the cooperating pairs, with respect to some specific criteria for the resources.

\begin{defi}\label{pairs}
Two different atoms $a$ and $b$ in $\phi$, with resources $z$ and $\tilde{z}$, respectively, are said to be in \textit{Mutually Nearest Neighbor Relation in $\mathbb{H}^3$ ($MNNR_{\mathbb{H}^3,\mathcal{D}}$)} if $a \stackrel{\phi}{\rightarrow} b$, $b \stackrel{\phi}{\rightarrow} a$, and if $(z,\tilde{z})\in \mathcal{D}$. We denote this by $a \stackrel{\phi,\mathcal{D}}{\leftrightarrow} b$. In telecommunication terms, the BSs $a$ and $b$ are in cooperation.
\end{defi}

\begin{defi}\label{singles}
An atom $a=(x,y,z)$ is said to be single if it is not in $MNNR_{\mathbb{H}^3,\mathcal{D}}$ (does not cooperate with any other atom in $\phi$). That is, if for every $b=(\tilde{x},\tilde{y},\tilde{z})$ in $\phi \backslash \{a\}$ such that $a \stackrel{\phi}{\rightarrow} b$, then $b \stackrel{\phi}{\nrightarrow} a$ or $(z,\tilde{z})\not \in \mathcal{D}$. 
\end{defi}
In the following, we give an example of the use of the set $\mathcal{D}$ to control the creation of cooperative pairs. Assume that, given a fixed position for the BSs, these have some users assigned. Quantify the available resources of each BS by its residual capacity, \textit{i.e.}, the remaining capacity after serving its assigned users. In this way, the resources being high  (\textbf{H}) is translated into a BS being assigned few users. Let B1 and B2 be two BSs, whose locations are close (in the 2-dimensional sense). Moreover, suppose that the resources of both are \textbf{H}. Therefore, B1 and B2 are hyperbollically close (their resources are balanced and their product is big, see the previous subsection). If an operator considers appropriate to make B1 and B2 cooperate, it is sufficient to consider the $MNNR_{\mathbb{H}^3,\mathcal{D}}$ with the control set $\mathcal{D}=(0,\infty)\times (0,\times \infty)$ (that is, without any extra constraint for the resources). On the other hand, suppose that another operator wants to minimize communication between the BSs (to prevent overburding the backhaul/control channel). Then, the operator might consider the cooperation between B1 and B2 unnecesary (the available resources of both are \textbf{H}, hence each one can serve sufficiently its own users). To block their cooperation, simply apply the $MNNR_{\mathbb{H}^3,\tilde{\mathcal{D}}}$ criterion, with a control set $\tilde{\mathcal{D}}$ chosen in an appropriate manner.  

\section{Analytic results for PPP}\label{SecIII}

Consider an homogeneous and independently marked PPP $\Phi$. It models the positions of the BSs over $\mathbb{R}^2$, with fixed density $\lambda>0$. The marks lie in $(0,\infty)$, they follow a common distribution $f(z)dz$, and represent the resources of each one of the BSs. This process turns out to be a PPP over $\mathbb{R}^2 \times (0,\infty)$ \cite{BacBlaVol1}, \textit{stationary with respect to the BS positions}, whose intensity measure is 
\begin{equation}\label{intensityM}
\Lambda(dx dy dz)=\lambda dx dy f(z) dz. 
\end{equation}
Notice that this measure is absolutely continuous with respect to the Lebesgue measure on $\mathbb{R}^2\times (0,\infty)$. From a different point of view, in $\mathbb{H}^3$, an hyperbolic volume measure $v(dxdydz)$ arises naturally from the hyperbolic metric  \cite{GroActHypSpa1998}, 
\begin{equation*}
v(dxdydz)=\frac{dxdydz}{z^3}.
\end{equation*}
If $\Lambda(\cdot)$ was absolutely continous with respect to $v(dxdydz)$, the resources of the BSs would be accumulated around the value $z=0$, $\mathbb{P}$-a.s., which is not realistic. However, for the interested reader, the whole analysis presented in this Section stays the same in both cases (just substitute $f(z)$ for $f^*(z)=\frac{f(z)}{z^3}$ in the formulas). 

\subsection{The probabililty of being in cooperation}

In this subsection, we fix $a$ and $b$ be two different atoms in $\mathbb{H}^3$, whose resources are $z$ and $\tilde{z}$, respectively. We fix also a control subset $\mathcal{D}$. Let 
\begin{equation*}
\begin{split}
R:=d_{\mathbb{H}^3}(a,b).
\end{split}
\end{equation*}
We consider the 3 dimensional Euclidean set 
\begin{equation}\label{setC}
C(a,b):=B_E(R,a)\cup B_E(R,b),
\end{equation}
(see Definition \ref{EucBall}, with $\epsilon=R$). 

Suppose that $a$ and $b$ belong to a configuration $\phi$. In geometric terms, the relation $a \stackrel{\phi}{\rightarrow} b$ holds if and only if (iff) the Hyperbolic ball $B_{\mathbb{H}^3}(R,a)$ is empty of atoms in $\phi \backslash \{a\}$. The latter happens iff the Euclidean ball $B_E(R,a)$ is empty of atoms in $\phi \backslash \{a\}$. Thus, the relation $a \stackrel{\phi,\mathcal{D}}{\leftrightarrow} b$ holds iff $C(a,b)$ is empty of atoms in $\phi\backslash \{a,b\}$, and if $(z,\tilde{z})\in \mathcal{D}$. Considering the empty space function of the PPP $\Phi$ \cite{BacBlaVol1}, we have the following result.

\begin{figure}[t!]
\centering
\includegraphics[trim = 0mm 0mm 0mm 0mm, clip, width=250pt]{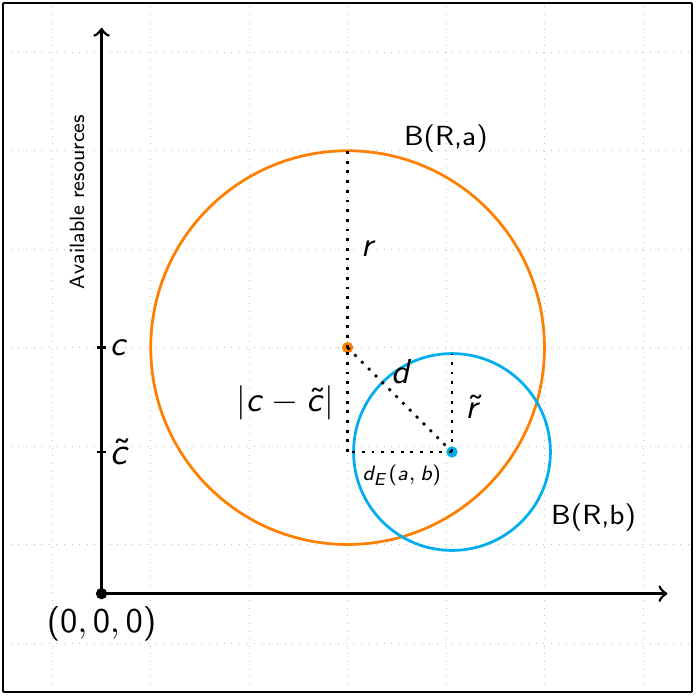}
\caption{The set $C(a,b)$, along with its representive variables.}
\label{Int1}
\end{figure}

\begin{prop}\label{prop1}
The probability of the atoms $a$ and $b$ being in MNNR is equal to 
\begin{equation}\label{videProba}
\mathbb{P}^{a,b}\left( a \stackrel{\Phi,\mathcal{D}}{\leftrightarrow}b \right)=e^{-\Lambda(C(a,b))}\mathbf{1}_{\{(z,\tilde{z})\in \mathcal{D}\}}, \ \mathbb{P}-a.s.,
\end{equation}
where the measure $\Lambda( \cdot)$ is given in equation \eqref{intensityM}, and $\mathbb{P}^{a,b}$ is the two fold Palm measure of $\Phi$.
\end{prop}
Compare the previous equation with the one in \cite[Lem. 1]{GioAnalInt2015}. Notice that they are practically the same. However, in \cite[Lem. 1]{GioAnalInt2015}, the two relevant balls are in $\mathbb{R}^2$, and their centers lie on the circumference of each other. Thus, the surface of its union can be easily calculated. In the model presented in this article, to find an analytic expression for the probability of two atoms being in MNNR$_{\mathbb{H}^3}$, we need to calculate the $\Lambda(\cdot)$-volume of the Euclidean set $C(a,b)$. At this point, we only know that $C(a,b)$ is the union of the 3-dimensional balls $B_E(R,a)$ and $B_E(R,b)$. To inquire into this task, let 
\begin{equation*}
\begin{split}
&(i) \ \ \ d:=\sqrt{d_E(a,b)^2+(z-\tilde{z})^2cosh^2(R)}, \\
&(ii) \ \ r:=zsinh(R), \ \ \ \ \tilde{r}:=\tilde{z}sinh(R), \\
&(iii) \ \, c:=zcosh(R), \ \ \ \, \tilde{c}:=\tilde{z}cosh(R).
\end{split}
\end{equation*}
Notice that $(i)$ is the 3-dimentional Euclidean distance between the centers of $B_E(R,a)$ and $B_E(R,b)$, $(ii)$ are the correspondig radii of $B_E(R,a)$ and of $B_E(R,b)$, and $(iii)$ are the corresponding third coordinate of $B_E(R,a)$ and $B_E(R,b)$ centers (see Figure \ref{Int1}). 
Define as well 
\begin{equation*}
\begin{split}
&h:=\frac{(\tilde{r}-r+d)(r+\tilde{r}-d)}{2d}, \ \ \tilde{h}:=\frac{(r-\tilde{r}+d)(r+\tilde{r}-d)}{2d}, \\
&\delta:=\frac{\pi}{2}-asin\left(\frac{\tilde{c}-c}{d}\right), \ \ \ \ \ \ \tilde{\delta}:=\frac{\pi}{2}-asin\left(\frac{c-\tilde{c}}{d}\right).
\end{split}
\end{equation*}
Since the variables $R$, $d$, $r$, $\tilde{r}$, $c$, $\tilde{c}$, $h$, $\tilde{h}$, $\delta$, $\tilde{\delta}$, are explicity given as functions of $d_E(a,b)$, $z$ and $\tilde{z}$, we have the following result.

\begin{theo}\label{theo1}
Let $F:(0,\infty)^3\longrightarrow (0,\infty)$ be the function, independent of the density $\lambda$ and $\mathcal{D}$, given by the expression 
\begin{equation}\label{balls}
\begin{split}
& F(d_E(a,b),z,\tilde{z}) = \\
& \lambda \pi \int \limits ^{r}_{-r}f(w+c)(r^2-w^2)dw +  \lambda \pi \int\limits^{r_2}_{-r_2}f(w+\tilde{c})(\tilde{r}^2-w^2)dw \\
&-\lambda \int\limits^{r}_{r-h}\int \limits^{2\pi}_0\int \limits^{\sqrt{r^2-w^2}}_0  f(c+wcos(\delta)-scos(\theta)sin(\delta)) s ds d\theta dw \\
& - \lambda \int \limits ^{\tilde{r}}_{\tilde{r}-\tilde{h}}\int \limits ^{2\pi}_0\int\limits ^{\sqrt{\tilde{r}^2-w^2}}_0 
f(\tilde{c}+wcos(\tilde{\delta})-scos(\theta)sin(\tilde{\delta})) s ds d\theta dw
\end{split}
\end{equation}
Then,
\begin{equation}\label{probaCon}
\mathbb{P}^{a,b}\left( a \stackrel{\Phi,\mathcal{D}}{\leftrightarrow} b \right) =e^{-\lambda F(d_E(a,b) ,z,\tilde{z})}\mathbf{1}_{\{(z,\tilde{z})\in \mathcal{D}\}}
\end{equation}
\end{theo}
We further provide a discussion on the above Theorem. The next Lemma reveals a little bit more on the topology of $C(a,b)$, defined in equation \eqref{setC}.

\begin{lem} \label{conex}
The 3 dimensional Euclidean balls $B(R,a)$ and $B(R,b)$ are never contained one within the other. Further, both balls always intersect each other, that is, $C(a,b)$ is connected.
\end{lem}
The previous Lemma and the inclusion-exclusion principle imply that 
\begin{equation}\label{AddSubs}
\begin{split}
\Lambda & (C(a,b)) \\ 
& = \Lambda(B(R,a))+\Lambda(B(R,b))-\Lambda(B(R,a)\cap B(R,b)).
\end{split}
\end{equation}
Then, the Lebesgue change of variable Theorem \cite[Th. 2.26]{RealComAnRud} allows us to give analytic expressions for $\Lambda(B(R,a))$, $\Lambda(B(R,b))$, and $\Lambda(B(R,b) \cap B(R,a))$, which are explicity given as functions of $R$, $d$, $r$, $\tilde{r}$, $c$, $\tilde{c}$, $h$, $\tilde{h}$, $\delta$, $\tilde{\delta}$. Following the above reasoning, and after substituing the analytic expression for $C(a,b)$ in equation \eqref{videProba}, we can prove Theorem \ref{theo1}.

\subsection{Interference analysis}
Via the dependent thinning defined in Section \ref{SecII}, we split $\Phi$ into two processes.

\begin{defi}
The process of singles and the process of cooperative pairs, $\Phi^{(1)}$ and $\Phi^{(2)}$, are given by 
\begin{equation*}
\begin{split}
& \Phi^{(1)}:=\{a\in \Phi \ | \ \mbox{ a is single } \} \\
&  \Phi^{(2)}:=\{a\in \Phi \ | \ \mbox{ a cooperates with another element of } \Phi \}
\end{split}
\end{equation*}
\end{defi}
 For two measurable functions $g:\mathbb{R}^2 \longrightarrow [0,\infty)$ and $k:\mathbb{R}^2 \times \mathbb{R}^2 \longrightarrow [0,\infty)$, we write the interference fields generated by the processes of singles and pairs by 
\begin{equation}\label{Int}
\begin{split}
\mathcal{I}^{(1)}_g & := \sum_{a \in \Phi^{(1)}} g(\hat{a}), \\ 
\mathcal{I}^{(2)}_k :&= \frac{1}{2}\sum_{a \in \Phi} \sum_{b \in \Phi  \backslash \{a\}} k(\hat{a},\hat{b}) 
\mathbf{1}_{\left\{a \stackrel{\Phi,\mathcal{D}}{\leftrightarrow} b \right\}}.
\end{split}
\end{equation}
The $1/2$ in front of the summation in \eqref{Int} prevents us from considering a pair twice. 

In this section, for every two $\hat{a},\hat{b}$ in $\mathbb{R}^2$, we denote by $d_E(\hat{a},\hat{b})$ the two-dimensional Euclidean distance between them.

Applying the Campbell-Little-Mecke formula, Slivnyak-Mecke Theorem \cite{BacBlaVol1}, and using the explicit expression provided by Theorem \ref{theo1}, we have the following result. 

\begin{theo} \label{ExpSin}
The expected value of the Interference generated by $\Phi^{(1)}$ and $\Phi^{(2)}$ is given by  
\begin{equation}\label{ExpectedV}
\begin{split}
& \mathbb{E} \left[ \mathcal{I}^{(1)}_g \right]  = 2\pi \lambda^2 \int \limits_{\mathbb{R}^2} \text{ \footnotesize $g(\hat{a})d\hat{a}$} \int \limits^\infty_0 
\left( \text{ \footnotesize $ 1-\mathbb{E} \left[e^{-\lambda F(s,Z,\tilde{Z})} \textbf{1}^{(Z,\tilde{Z})}_\mathcal{D}\right] $} \right)
sds, \\
& \mathbb{E} \Bigg[ \mathcal{I}^{(2)}_k \Bigg] = \frac{\lambda^2}{2} \int\limits_{\mathbb{R}^2} \int\limits_{\mathbb{R}^2} \text{ \footnotesize $ k(\hat{a},\hat{b}) \mathbb{E}\left[e^{-\lambda F(d_E(\hat{a},\hat{b}),Z,\tilde{Z})}\textbf{1}^{(Z,\tilde{Z})}_\mathcal{D}\right]$}d \hat{a} d\hat{b},
\end{split}
\end{equation} 
where $Z$ and $\tilde{Z}$ are two independent random variables, with common distribution $f(z)dz$, and $F(s,z,\tilde{z})$ is defined in Theorem \ref{theo1}, equation \eqref{balls}.
\end{theo}
Notice that, to calculate the expected value of the interference generated by the process of singles, we need to compute the integral of $g(\cdot)$ over $\mathbb{R}^2$, and then multiply it by another integral. However, to calculate the expected value of the interference generated by the process of pairs, we have to compute an integral over $\mathbb{R}^2 \times \mathbb{R}^2$ of the function $k(\cdot,\cdot)$ times another function that performs as a proportion of two points being in a cooperative pair.     

\subsection{Percentage of atoms in cooperative pairs}
Equation \eqref{probaCon} is an expression for the probability of two given atoms being in cooperation. This expression is in function of their Euclidean distance and their corresponding available resources. Therefore, we can interpret Theorem \ref{theo1} as a local result. We would like to go further, and give a global metric for the atoms in cooperative pairs, which does not depend on the position.

Let $M^{(1)},M^{(2)}:\mathcal{B}(\mathbb{R}^2) \longrightarrow [0,\infty)$ be the measures such that 
\begin{align*}
M^{(1)}(A) & =\mathbb{E}\Bigg[ \sum_{a\in \Phi^{(1)}}\mathbf{1}_{\{ \hat {a}\in A \} } \Bigg], \\ 
M^{(2)}(A) & =\mathbb{E}\Bigg[ \sum_{a\in \Phi^{(2)}}\mathbf{1}_{\{ \hat {a}\in A \} } \Bigg], 
\end{align*}
for every $A\in \mathcal{B}(\mathbb{R}^2)$. The number $M^{(1)}(A)$ (respectively $M^{(2)}(A)$) gives the average number of single (respectively cooperating) BSs, whose positions lie inside $A$. 

To give an expression for the intensity measure of the two processes, it is possible to use the formulas in Theorem \ref{ExpSin}, as follows: Fix $A\in \mathcal{B}(\mathbb{R}^2)$ and, for every $a$ and $b$ in $\mathbb{H}^3$, consider the function $k(\hat{a},\hat{b}):=2 \mathbf{1}^{\hat{a}}_A$. Hence, 
\begin{equation*}
\begin{split}
\mathcal{I}^{(2)}_k  & = \frac{1}{2} \sum_{a\in \Phi}\sum_{b\in \Phi\backslash \{a\}} k(\hat{a},\hat{b}) \mathbf{1}_{\{a \stackrel{\Phi,\mathcal{D}}{\leftrightarrow} b \}}\\
& = \sum_{a\in \Phi}  \mathbf{1}^{\hat{a}}_A \sum_{b\in \Phi \backslash \{a\}} \mathbf{1}_{\{a \stackrel{\Phi,\mathcal{D}}{\leftrightarrow} b \}}, \ \mathbb{P}-a.s.  
\end{split}
\end{equation*}
Since the nearest neighbor is unique $\mathbb{P}$-a.s., then, for every atom $a\in \Phi$,
\begin{equation*}
\sum_{b\in \Phi \backslash \{a\}} \mathbf{1}_{\{a \stackrel{\Phi,\mathcal{D}}{\leftrightarrow} b \}} = \mathbf{1}_{\{ a\in \Phi^{(2)}\} }, \ \mathbb{P}-a.s.
\end{equation*}
Therefore, 
\begin{equation*}
\mathcal{I}^{(2)}_k   = \sum_{a\in \Phi^{(2)}}  \mathbf{1}_{ \{ \hat{a}\in A\}}, \ \mathbb{P}-a.s.
\end{equation*}
Taking the expected value on the previous equation, we have that, for this particular choice of $k(\hat{a},\hat{b})$,
\begin{equation}\label{M22}
M^{(2)}(A)=\mathbb{E}\left[ \mathcal{I}^{(2)}_k\right].
\end{equation}

\begin{theo} \label{proporDoub}
There exists a number $P_{\mathcal{D}}(\lambda,f)\in [0,1]$, depending on $\lambda$, the density function $f(\cdot)$, and the subset $\mathcal{D}$, such that, for every $A\in \mathcal{B}(\mathbb{R}^2)$,
\begin{align*}
M^{(1)}(A) & = (1-P_\mathcal{D}(\lambda,f))\lambda \mathcal{S}(A), \\
M^{(2)}(A) & =P_{\mathcal{D}}(\lambda,f) \lambda \mathcal{S}(A). 
\end{align*}
Furthermore, 
\begin{equation} \label{numberP}
P_{\mathcal{D}}(\lambda,f)= 2 \pi \lambda \int\limits^\infty_0\mathbb{E} \left[ e^{-\lambda F(s,Z,\tilde{Z})} \mathbf{1}^{(Z,\tilde{Z})}_\mathcal{D}\right] s ds,
\end{equation}
where $F(s,z,\tilde{z})$ is defined y Theorem \ref{theo1}, and $Z$ and $\tilde{Z}$ are two independent random variables, with common distribution $f(z)dz$. 
\end{theo}
Notice that the number $P_{\mathcal{D}}(\lambda,f)$ does not depend on the surface $A$. The BS positions are modeled by a PPP, with intensity $\lambda$. Then, the previous Theorem states that the intensity of singles and cooperative pairs among the BSs is 
\begin{equation*}
\begin{split}
& (1-P_\mathcal{D}(\lambda,f))\lambda,\ P_\mathcal{D}(\lambda,f)\lambda,
\end{split}
\end{equation*} 
respectively. Thus, we can interpret $(1-P_\mathcal{D}(\lambda,f))$ and $P_\mathcal{D}(\lambda,f)$ as a \textit{the proportion of singles and cooperative pairs}.

For numerical evaluation, notice that $P_\mathcal{D}(\lambda,f)$ can be evaluated either $(i)$ via Monte Carlo simulation, since
\begin{equation*}
\begin{split}
 \mathbb{E} \Big[ \textbf{1}^{(Z,\tilde{Z})}_\mathcal{D} & \int^\infty_0 e^{-\lambda F(r,Z,\tilde{Z})} r dr \Big]  \\
& \approx \sum^N_{=1i}  \frac{\textbf{1}^{(Z_i,\tilde{Z}_i)}_\mathcal{D} \int^\infty_0 e^{-\lambda F(r,Z_i,\tilde{Z}_i)}}{N},
\end{split}
\end{equation*}
(where $(Z_i,\tilde{Z}_i)^N_{i=1}$ is an independent family of random vectors, with common distribution $f(z)dz f(\tilde{z})d\tilde{z}$), or $(ii)$ via numerical integration, since 
\begin{equation*}
\begin{split}
\mathbb{E} \Big[ \textbf{1}^{(Z,\tilde{Z})}_\mathcal{D} & \int^\infty_0 e^{-\lambda F(r,Z,\tilde{Z})} r dr \Big]  \\
& =  \iint_\mathcal{D} \left(  \int^\infty_0 e^{-\lambda F(r,z,\tilde{z})} r dr\right)f(z)f(\tilde{z})dzd\tilde{z}.
\end{split}
\end{equation*}


\section{Numerical Evaluation and Examples}\label{SecIV}
In this section, we consider a density for the BSs $\lambda=1 \ [km^2]$.

\subsection{Proportion of Single BSs and Cooperative Pairs}
\begin{figure}[t!]
\center
\includegraphics[trim = 10mm 60mm 20mm 60mm, clip, width=240pt]{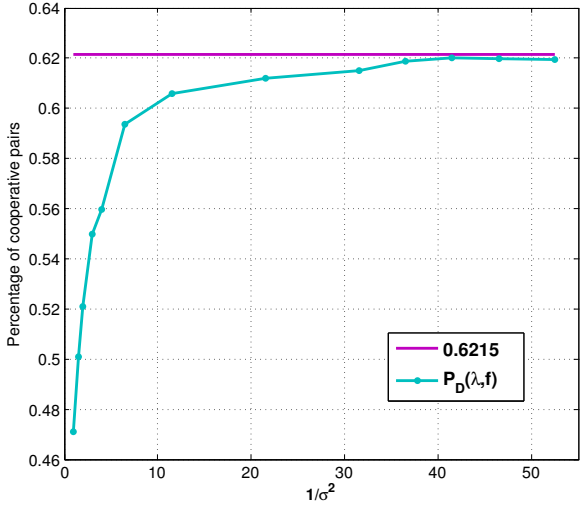}
\caption{Percentage of cooperative pairs. The marks are Beta distrubuted, centered at $0.5$, with different values for the variance $\sigma^2$.}
\label{PercDoub}
\end{figure}

For the following evaluations, we assume no extra constraint for the marks, that is, we choose $\mathcal{D}=(0,\infty) \times (0,\infty)$. From equations \eqref{balls}, \eqref{probaCon}, and \eqref{numberP}, notice that the proportion of cooperative pairs depends strongly on the distribution of the marks. In Figure \ref{PercDoub} we provide the numerical evaluation of $P_\mathcal{D}(\lambda,f)$. Specifically, in this example, the marks are distributed as a Beta random variable, with mean value $\mu=0.5$. Recall that the Beta distribution is defined by its two first moments. The figure illustrates how \textit{the percentage of stations in pair varies with the change of the variance of the marks}. Observe that, when the variance goes to zero, the value of $P_\mathcal{D}(\lambda,f)$ tends to $0.6215$, which is the average number of cooperative pairs in the strictly geometric model of \cite{GioAnalInt2015,CovGainsAlv2016}. On the other hand, when the variance is large, the value of $P_\mathcal{D}(\lambda,f)$ differs from  $0.6215$ significantly. 

To understand this better, suppose that $\mu>0$ is the mean value of the marks. If the variance is large, the available resources oscillate considerably around $\mu$. Equation \eqref{hyperDist} tells us that both the term involving the load balancing and the product of the available resources play a role in the formation of the cooperative pairs (instead of merely the Euclidean proximity between the BS positions, as in \cite{GioAnalInt2015,CovGainsAlv2016}). On the other hand, for cellular networks for which the available resources of the BSs do not vary considerably around $\mu$ (that is, the resources are uniformly available throughout the network), the load balancing term and the product of their available resources are practically constant. 
Then, from equation \eqref{hyperDist} it is clear that the 2-dimensional Euclidean distance between the BSs would be the most influential. In this case, we practically recover the model in \cite{GioAnalInt2015,CovGainsAlv2016}.

\subsection{Expected value of the interference field}
Since we are under a stationary framework, we suppose that the typical user is placed at the Euclidean origin $\hat{0}:=(0,0)$.

\begin{figure}[htbp] 
\center
\includegraphics[trim = 10mm 70mm 10mm 70mm, clip, width=250pt]{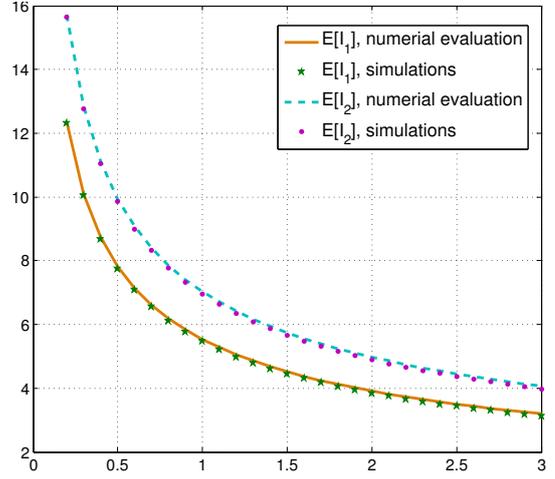}
\caption{Expected value of the interference flied generated by the singles and pairs, for $\beta=2.5$. Numerical evaluations and simulations results.}
\label{Exp}
\end{figure}

Given the position $\hat{a}$ of a single BS, we suppose that it transmits a signal/interference $s(\hat{a})$, towards the typical user. For example, given a pathloss exponent $\beta>2$, we can take $s(\hat{a}):=\frac{1}{(d_E(\hat{a},\hat{0}))^\beta}$. Consider in equation \eqref{Int} the function $g(\hat{a}):=s(\hat{a})\textbf{1}_{\{ d_E(\hat{a},\hat{0})>R\} }$, where $R>0$ is fixed. The indicator function serves to calculate the interference generated by the single BSs whose distance to the typical user is larger than $R$. The numerical evaluation of $\mathbb{E}\left[\mathcal{I}^{(1)}_g\right]$, using Theorem \ref{ExpSin}, is given in Figure \ref{Exp}. Notice that the expression in equation \eqref{ExpectedV} gives almost identical results with the simulations. 

To calculate the interference from pairs, we make the following choice of the function $k(\cdot,\cdot)$. For two BSs at 2-dimensional positions $\hat{a}$ and $\hat{b}$, that form a cooperative pair, suppose that these transmit orthogonal signals, which are added at the typical user: $s(\hat{a})+s(\hat{b})$. Given a pathloss exponent $\beta>2$, we can take $s(\hat{a}):=\frac{1}{(d_E(\hat{a},\hat{0}))^\beta}$. Then, consider in equation \eqref{Int} the function $$k(\hat{a},\hat{b}):=s(\hat{a})\textbf{1}_{\{ d_E(\hat{a},\hat{0})>R\} }+s(\hat{b})\textbf{1}_{\{ d_E(\hat{b},\hat{0})>R\} },$$ where $R>0$ is fixed. After some calculations, 
\begin{equation*}
\mathbb{E}[\mathcal{I}^{(2)}]=2\pi \lambda^2 \int\limits_{d(\hat{c},\hat{0})>R} s(\hat{c})d\hat{c} \int\limits^\infty_0 \mathbb{E}\left[e^{-\lambda \tilde{F}(s,Z,\tilde{Z})}\textbf{1}^{(Z,\tilde{Z})}_\mathcal{D}\right]s ds.
\end{equation*}  
The numerical evaluation of $\mathbb{E}\left[ \mathcal{I}^{(2)}_k\right]$ is given in Figure \ref{Exp}. Again, the expression in equation \eqref{ExpectedV} gives, as well, almost identical results with the simulations. 

\section{Conclusions}\label{SecV}

In this paper, we proposed a novel grouping criterion of BSs, which extends the MNNR criterion introduced by the authors in \cite{GioAnalInt2015,CovGainsAlv2016}. The resulting clusters are single BSs and cooperative pairs. To form the cooperative pairs, the criterion favors the BSs which are geographically close, and, at the same time, have both enough resources, so that their cooperation is meaningful and beneficial for the network. When the BSs are modeled by a PPP, an analysis of the probability of two BSs being in a cooperative pair, followed by an interference analysis, are provided. In particular, for cellular networks where the available resources of the BSs vary a lot, the percentage of formed pairs differs considerably from the purely geometrically model in \cite{GioAnalInt2015,CovGainsAlv2016}. On the other hand, for cellular networks where the available resources of the BSs stay almost constant throughout the network, the average number of BSs in cooperation is, as expected, close to that one in \cite{GioAnalInt2015,CovGainsAlv2016}.

\section{Appendix}

\subsection{The height of a lens}\label{heightH}
For this subsection, capital letters denote points in the Euclidean space, and small letters denote length of segments. 

\begin{figure}[t!]
\center
\includegraphics[trim = 0mm 0mm 0mm 0mm, clip, width=200pt]{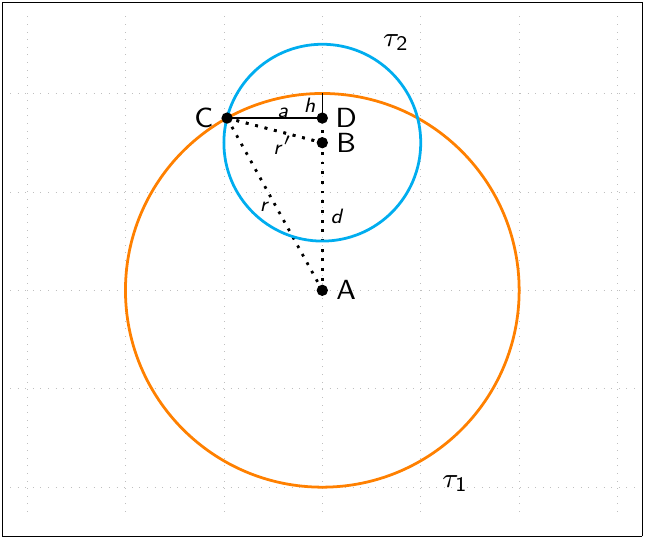}
\caption{The angle $\delta$ and the subsets $I_1$ and $I_2$.}
\label{IntPreuve}
\end{figure}

Let $\tau_1$ and $\tau_2$ be two Euclidean spheres, both intersecting each other (see Figure \ref{IntPreuve}). Suppose that their respective centers, $A$ and $B$, and their respective radii, $r$ and $r'$, are known. In particular, the value of $d$ (the length of the segment $\overline{AB}$) is known as well. Denote by $h$ the length of the upper height of the lens described by the Intersection of $\tau_1$ and $\tau_2$ (again, see Figure \ref{IntPreuve}). The aim of this subsection is to give an expression for $h$ in function of $r$, $r'$, and $d$. 

Denote by $\triangle ABC$ the 2-dimensional Euclidean triangle defined by $A$, $B$, and $C$, and by $(ABC)$ its 2-dimensional Euclidean surface. Observe that the segment $\overline{CD}$ is the height of the triangle $\triangle ABC$. Then, if $a$ denotes the length of $\overline{CD}$, $(ABC)=\frac{a.d}{2}$ and, in particular, $a=\frac{2(ABC)}{d}$. Applying the Phytagorean Theorem to the triangle $\triangle ACD$, we have that $(r-h)^2+a^2=r^2$, and therefore, $h=r-\sqrt{r^2-a^2}$.

On the other hand, Heron's formula \cite{IntGeoCox} gives an expression for $(ABC)$ in function of $r$, $r'$ and $d$,
\begin{equation*}
(ABC)=\frac{\sqrt{2r^2(r')^2+2r^2d^2+2(r')^2d^2-r^4-(r')^4-d^4}}{4}.
\end{equation*}
 Since 
\begin{equation*}
\begin{split}
r^2&-a^2 = r^2+ \frac{4(ABC)^2}{d^2} \\
&= r^2+\frac{-2r^2(r')^2-2r^2d^2-2(r')^2d^2+r^4+(r')^4+d^4}{4d^2} \\
&=\frac{-2r^2(r')^2-2(r')^2d^2+2r^2d^2+r^4+(r')^4+d^4}{4d^2} \\
&=\frac{(r^2-(r')^2+d^2)^2}{4d^2}, 
\end{split}
\end{equation*}
thus, 
\begin{equation*}
\begin{split}
h&=r-\sqrt{r^2-a^2} \\
&=r-\frac{r^2-(r')^2+d^2}{2d} \\
&=\frac{(r'-r+d)(r'+r-d)}{2d}.
\end{split}
\end{equation*}

\subsection{Proof of Proposition \ref{prop1}.}
Fix $\epsilon>0$, and an atom $a=(x,y,z)$ in $\mathbb{H}^3$. Another atom $b=(\tilde{x},\tilde{y},\tilde{z})$ belongs to the hyperbolic ball $B_{\mathbb{H}^3}(\epsilon,a)$ iff 
\begin{equation*}
 acosh\left(\frac{d_E(a,b)^2}{2z \tilde{z}} + \frac{1}{2}\left( \frac{z}{\tilde{z}}+\frac{\tilde{z}}{z}\right) \right) < \epsilon .
\end{equation*}
Then, after some manipulations, we have that the atom $b$ belongs to $B_{\mathbb{H}^3}(\epsilon,a)$ iff 
\begin{equation*}
d_E(a,b)^2 + \tilde{z}^2 -2z \tilde{z}cosh(\epsilon)+z^2 < 0.
\end{equation*}
Remark that 
\begin{equation*}
\begin{split}
z^2 - & 2z \tilde{z}cosh(\epsilon)+\tilde{z}^2  \\
&= z^2 -z^2cosh^2(\epsilon)+z^2cosh^2(\epsilon)-2z \tilde{z}cosh(\epsilon)+\tilde{z}^2 \\
& = z^2(1-cosh^2(\epsilon))+(zcosh(\epsilon)-\tilde{z})^2 \\
& = -z^2sinh^2(\epsilon)+(zcosh(\epsilon)-\tilde{z})^2 ,
\end{split}
\end{equation*}
where the last equality holds after considering the hyperbolic trigonometric identity $cosh^2(\epsilon)-1=sinh^2(\epsilon)$. Recall that $d_E(a,b)^2=(x-\tilde{x})^2+(y-\tilde{y})^2$, then we conclude that the atom $b=(\tilde{x},\tilde{y},\tilde{z})$ belongs to $B_{\mathbb{H}^3}(\epsilon,a)$ iff $$(x-\tilde{x})^2+(y-\tilde{y})^2+(zcosh(\epsilon)\tilde{z})^2 <z^2sinh^2(\epsilon).$$ Since $a=(x,y,z)$ is fixed, the previous equation describes a ball in the Euclidean space $\mathbb{R}^3$, with center at $(x,y,zcosh(\epsilon))$, and radius $zsinh(\epsilon)$. 

\subsection{Uniqueness of the nearest neighbor}
Let $\Phi$ be a point process stationary point process on $\mathbb{R}^2$, with density $0<\lambda<\infty$, independently marked over $(0,\infty)$, and whose marks have $f(z)dz$ as their common distribution. Its intensity measure $\Lambda(\cdot)$ has the same representation as in equation \eqref{intensityM} \cite{BacBlaVol1}. Consider the mapping 
\begin{equation*}
\begin{split}
\mathbb{H}^3 \times \mathbb{H}^3 & \rightarrow \mathbb{H}^3 \\
                                                   (b,a) & \mapsto      (b\ominus a)
\end{split}
\end{equation*}
such that, for every two atoms $a=(x,y,z)$ and $b=(\tilde{x},\tilde{y},\tilde{z})$, $(b\ominus a)=(\tilde{x}-x,\tilde{y}-y,\tilde{z})$. Notice that this mapping is not symetric. Fixed $a\in \mathbb{H}^3$, for every simple and locally finite configuration $\phi=\{b_n\}^\infty_{n=1}$, we consider the following notation as in \cite{BacBlaVol1}
\begin{equation}\label{diff}
\phi-a:=\{b_n\ominus a\}^\infty_{n=1}, 
\end{equation}
For a typical element $c\in \mathbb{H}^3$, we will prove that there are no atoms from $\Phi$ equidistant to $c$, $\mathbb{P}$-a.s. Let 
\begin{equation*}
\mathcal{B}: =\{ \mbox{For some } a,b\in \Phi, \ a\neq b, \ d_{\mathbb{H}^3}(a,c)=d_{\mathbb{H}^3}(b,c)  \} 
\end{equation*}
To prove our point, we have to prove that $\mathbb{P}(\mathcal{B})=0$. Since the process is stationarity, we can suppose that $c=(0,0,z^*)$.
For every two atoms $a=(x,y,z)$ and $b=(\tilde{x},\tilde{y},\tilde{z})$, for this particular choice of $c$, it holds that  
\begin{equation*}
\begin{split}
d_{\mathbb{H}^3}(a,c) & = \frac{x^2+y^2}{2zz^*}+\frac{1}{2}\left( \frac{z}{z^*} +\frac{z^*}{z} \right) \\
d_{\mathbb{H}^3}(b,c) & = \frac{\tilde{x}^2+\tilde{y}^2}{2\tilde{z}z^*}+\frac{1}{2}\left( \frac{z^*}{\tilde{z}} +\frac{\tilde{ z}}{z^*} \right) \\
                                       & =\frac{ (\tilde{x}-x+x)^2+(\tilde{y}-y+y)^2}{2\tilde{z}z^*} +\frac{1}{2}\left( \frac{z^*}{\tilde{z}} +\frac{\tilde{ z}}{z^*} \right)
\end{split}
\end{equation*}
Then, $d_{\mathbb{H}^3}(a,c)=d_{\mathbb{H}^3}(b,c)$ happens iff
\begin{equation*}
\begin{split}
& \frac{x^2+y^2}{z}-\frac{ (\tilde{x}-x+x)^2+(\tilde{y}-y+y)^2}{\tilde{z}} \\
& = z^*\left( \frac{z}{z^*} +\frac{z^*}{z} - \left(\frac{z^*}{\tilde{z}} +\frac{\tilde{ z}}{z^*}\right) \right)
\end{split}
\end{equation*}
Let $\alpha,\beta:\mathbb{H}^3 \times \mathbb{H}^3 \rightarrow \mathbb{H}^3$ be the functions such that, for every pair of atoms $a=(x,y,z)$ and $b=(\tilde{x},\tilde{y},\tilde{z})$,
\begin{equation}\label{alpbet}
\begin{split}
\alpha(a,b) & :=\frac{x^2+y^2}{z}- \frac{ (\tilde{x}+x)^2+(\tilde{y}+y)^2}{\tilde{z}}, \\
\beta(a,b) & :=z^*\left( \frac{z^*}{\tilde{z}} +\frac{\tilde{ z}}{z^*} - \frac{z}{z^*} - \frac{z^*}{z}  \right).
\end{split}
\end{equation}
Then, for every two atoms $a$ and $b$, $d_{\mathbb{H}^3}(a,c)=d_{\mathbb{H}^3}(b,c)$ happens iff $\alpha(a,b\ominus a)-\beta(a,b)=0$. Therefore, we can rewrite $\mathcal{B}$ as follows
\begin{equation*}
\mathcal{B} =\{ \mbox{For some } a,b\in \Phi, \ a\neq b, \ \alpha(a,b\ominus a)-\beta(a,b)=0  \}.
\end{equation*}
For every $a\in \mathbb{H}^3$ and every simple, and locally finite cofiguration $\phi$ on $\mathbb{H}^3$, let
\begin{equation}\label{g1}
g(a,\phi):=\sum_{b\in \phi } \mathbf{1} _{\{ \hat{b}\neq 0, \  \alpha(a,b)-\beta(a,b)=0  \}}.
\end{equation}
In particular, 
\begin{equation}\label{g2}
\begin{split}
g(a,\phi-a) &=\sum_{b\in \phi-a } \mathbf{1} _{\{ \hat{b}\neq 0, \  \alpha(a,b)-\beta(a,b)=0  \}} \\
&\stackrel{(a)}{=}\sum_{b\in \phi} \mathbf{1} _{\{ \hat{b}\neq 0, \  \alpha(a,b\ominus a)-\beta(a,b)=0  \}},
\end{split}
\end{equation}
where $(a)$ follows from equation \eqref{diff}, and from the fact that the function $\beta(\cdot,\cdot)$ only depends on the resources of the atoms, not on their position. On the other hand, notice that
\begin{equation*}
\begin{split}
\mathcal{B} &=\bigcup_{a\in \Phi } \bigcup_{b\in \Phi } \{ b\neq a, \ \alpha(a,b\ominus a)-\beta(a,b)=0  \}.
\end{split}
\end{equation*}
Denote by $\mathbb{E}^0$ the Palm measure of $\Phi$. Hence, 
\begin{equation*}
\begin{split}
\mathbb{P}(\mathcal{B}) & \leq \mathbb{E}\Bigg[ \sum_{a\in \Phi } \sum_{b\in \Phi } \mathbf{1} _{\{ \hat{(b\ominus a)}\neq 0, \  \alpha(a,b\ominus a)-\beta(a,b)=0  \}} \Bigg] \\
& \stackrel{(a)}{=} \mathbb{E} \Bigg[ \int_{\mathbb{H}^3 } g(a,\Phi-a) \Phi(da) \Bigg] \\
& \stackrel{(b)}{=} \int_{\mathbb{H}^3 } \mathbb{E}^0 \left[ g(a,\Phi) \right] \Lambda(da) \\
& \stackrel{(c)}{=} \int_{\mathbb{H}^3 } \mathbb{E}^0 \left[ \sum_{b\in \Phi } \mathbf{1} _{\{ \hat{b}\neq 0, \  \alpha(a,b)-\beta(a,b)=0  \}} \right] \Lambda(da) \\
& \stackrel{(d)}{=}  \mathbb{E}^0 \left[ \sum_{b\in \Phi } \int_{\mathbb{H}^3} \mathbf{1} _{\{ \hat{b}\neq 0, \  \alpha(a,b)-\beta(a,b)=0  \}} \Lambda(da) \right],
\end{split}
\end{equation*}
where $(a)$ follows from equation \eqref{g2}, $(b)$ holds after considering the reduced Campbell's formula for an marked and stationary point process \cite{BacBlaVol1}, $(c)$ follows from equation \eqref{g1}, and $(d)$ after considering Tonelli's Theorem. 

Fix $b=(\tilde{x},\tilde{y},\tilde{z})$, recall that $z^*>0$ is fixed as well. From equation \eqref{alpbet}, notice that the transformation $a \mapsto \alpha(a,b) - \beta(a,b)=0$ defines a 2-dimensional surface in $\mathbb{H}^3$. Since $\Lambda(\cdot)$ is absolutely continous with respect to the Lebesgue measure, then
\begin{equation*}
\begin{split}
0 & = \Lambda(\{a \ | \ \alpha(a,b) - \beta(a,b)=0\}) \\
&= \int_{\mathbb{H}^3} \mathbf{1} _{\{ \hat{b}\neq 0, \  \alpha(a,b)-\beta(a,b)=0  \}} \Lambda(da).
\end{split}
\end{equation*}
We conclude that $\mathbb{P}(\mathcal{B})=0$.
\subsection{Proof of Lemma \ref{conex}}
The Euclidean ball $B(R,a)$ is contained within $B(R,b)$ iff $d+r<\tilde{r},$ (see Figure \ref{Int1}), which happens iff, $$\sqrt{d_E(a,b)^2+(z-\tilde{z})^2cosh^2(R)}<(\tilde{z}-z)sinh(R).$$ Since $cosh(R)\geq sinh(R)$, last inequality never holds, whether $z>\tilde{z}$ or $\tilde{z}>z$. This implies that $B(R,a)$ is never contained inside $B(R,b)$. By symmetry, $B(R,b)$ is never contained within $B(R,a)$ either.

The balls $B(R,a)$ and $B(R,b)$ intersect each other iff $r+\tilde{r}>d$ (again, see Figure \ref{Int1}). Square the previous inequality, then consider the values of $r$, $\tilde{r}$, $d$, and again use the identity $cosh^2(\epsilon)-1=sinh^2(\epsilon)$. Then, the balls $B(R,a)$ and $B(R,b)$ intersect each other iff
\begin{equation*}
 (z+\tilde{z})^2(cosh^2(R)-1)>d_E(a,b)^2+(z-\tilde{z})^2cosh^2(R).
\end{equation*}
Notice that, for genera values ofl $z,\tilde{z},R,d_E(a,b)>0$, this inequality does not always hold (choose small values of $z$, $\tilde{z}$, and $R$, and a large value of $d_E(a,b)$). In our particular situation, $R$ is the hyperbolic distance between $a$ and $b$ (see equation \ref{hyperDist}). Hence, $r+\tilde{r}>d$ holds iff, 
\begin{equation*}
\begin{split}
& (z+\tilde{z})^2\left(\left(\frac{d_E(a,b)^2}{2z\tilde{z}}+\frac{1}{2}\left(\frac{z}{\tilde{z}}+\frac{z}{\tilde{z}}\right) \right)^2-1\right) \\
& \ \ \ \ \ \ \ \ >d_E(a,b)^2+(z-\tilde{z})^2\left(\frac{d_E(a,b)^2}{2z\tilde{z}}+\frac{1}{2}\left(\frac{z}{\tilde{z}}+\frac{\tilde{z}}{z}\right) \right)^2  
\end{split}
\end{equation*}
After some simple manipulations, we verify that the previous inequality is equivalent to 
\begin{equation*}
\begin{split}
3\left(z-\tilde{z}\right)^2&+\frac{d_E(a,b)^2}{2z\tilde{z}} \\
& +d_E(a,b)^2 \left(2\left(\frac{z}{\tilde{z}}+\frac{\tilde{z}}{z}\right)-1\right) +\frac{(z-\tilde{z})^4}{z\tilde{z}}>0.
\end{split}
\end{equation*}
Since $2\left(\frac{z}{\tilde{z}}+\frac{\tilde{z}}{z}\right)>2$, the preceding inequality always holds.


\subsection{Proof of Theorem \ref{theo1}}
To make more clear the proof, we provide to the reader this table, containing the description of the variables 
\begin{center} \label{table1}
    \begin{tabular}{ | c |  p{5.5cm} |}
    \hline
    Symbol                                & Description                    \\ \hline
    $a$                                      & Atom in $\mathbb{H}^3$, with available resources $z>0$   \\ \hline
    $b$                                      & Atom in $\mathbb{H}^3$, with available resources $\tilde{z}>0$   \\ \hline
    $R$                                      & $d_{\mathbb{H}^3}(a,b)$, the hyperbolic distance between $a$ and $b$ \\ \hline
    $d_E(a,b)$                          & The 2-dimensional Euclidean distance between the positions of $a$ and $b$             \\ \hline
    $B_{\mathbb{H}^3}(R,a)$ & The hyperbolic ball, centered at $a$, with radius $R$\\  \hline
    $B_{\mathbb{H}^3}(R,b)$ & The hyperbolic ball, centered at $b$, with radius $R$\\  \hline    
    $B_E(R,a)$                          & The 3-dimensional Euclidean ball described by $B_{\mathbb{H}^3}(R,a)$\\  \hline
    $B_E(R,b)$                          & The 3-dimensional Euclidean ball described by $B_{\mathbb{H}^3}(R,a)$\\  \hline
    $d$                                      & The distance between $B_E(R,a)$ and $B_E(R,b)$ centers \\  \hline   
    $r$                                       & The radius of $B_E(R,a)$  \\  \hline   
    $\tilde{r}$                            & The radius of $B_E(R,b)$  \\  \hline   
    $c$                                       & The third coordinate of $B_E(R,a)$ center  \\  \hline   
    $\tilde{c}$                           & The third coordinate of $B_E(R,b)$ center  \\  \hline   
    $C(a,b)$                              & The 3-dimensional Euclidean set, representing the union of $B_E(R,a)$ and $B_E(R,b)$  \\ \hline
 \end{tabular}
\end{center}
To obtain $C(a,b)$, we need to calculate $\Lambda(B(R,a))$, $\Lambda(B(R,b))$, and $\Lambda(B(R,a) \cap B(R,a))$ (see equation \eqref{AddSubs}). 

\begin{figure}[t!]
\center
\includegraphics[trim = 0mm 0mm 0mm 0mm, clip, width=200pt]{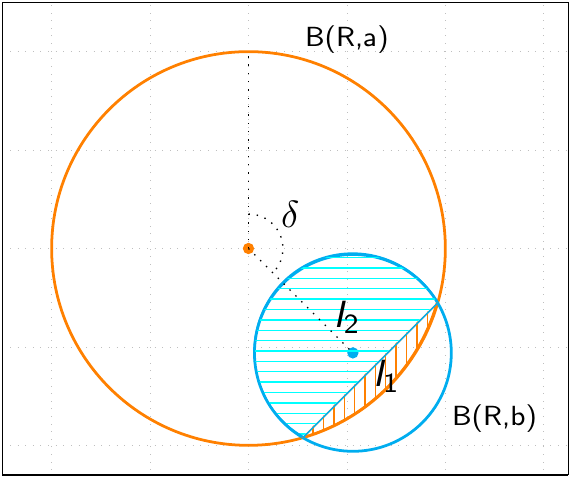}
\caption{The angle $\delta$ and the subsets $I_1$ and $I_2$.}
\label{Int2}
\end{figure}

The measure $\Lambda(\cdot)$ is invariant under translations and rigid transformations, with respect to the $xy$-axis. Thus, we can suppose that $a=(0,0,z)$, in particular, $B(R,a)$ is centered at $(0,0,c)$ (see Definition \ref{EucBall}).  Consider $\tau:\mathbb{R}^3 \rightarrow \mathbb{R}^3$, the translation sending $(0,0,c)$ to the origin. Then, $\tau(B(R,a))$ is a 3-dimensional Euclidean ball, centered at the origin, with radius $r$. Given that $|D\tau^{(-1)}|=1$ (the absolute value of the Jacobian), the Lebesgue change of variable Theorem \cite[Th. 2.26]{RealComAnRud} states that 
\begin{equation*}
\begin{split}
\Lambda(B(R,b)) 
&= \iiint\limits_{B(R,a)} \lambda f(w) dxdydw \\
&= \iiint\limits_{\tau(B(R,a))} \lambda f(w+c) dxdydw.
\end{split}
\end{equation*}
Observe that $\tau(B(R,a))$ can be described in cylindrical coordinates. Thus,
\begin{equation*}
\begin{split}
\Lambda(B(R,a)) &= \lambda \int\limits^r_{-r} \int\limits^{2\pi}_0 \int\limits^{\sqrt{r^2-w^2}}_0 f(w+c) s dsd\theta dw \\
&= \lambda 2 \pi \int\limits^r_{-\tilde{r}}f(w+c) \int\limits^{\sqrt{r^2-w^2}}_0  s ds  dw \\
&= \lambda \pi \int\limits^r_{-r}f(w+c) (r^2-w^2) dw.
\end{split}
\end{equation*}
In the same fashion, 
\begin{equation*}
\Lambda(B(R,b))= \lambda \pi \int\limits^{\tilde{r}}_{-\tilde{r}}f(w+\tilde{c}) (\tilde{r}^2-w^2) dw.
\end{equation*}
We only have left to calculate $\Lambda(B(R,a) \cap B(R,b))$. Again, we can consider $a=(0,0,z)$ and $b=(d_E(a,b),0,\tilde{z})$. In particular, $B(R,a)$ and $B(R,b)$ are centered at $(0,0,c)$ and $(d_E(a,b),0,\tilde{c}))$, respectively (again, see Definition \ref{EucBall}). Denote by $I_1$  ( $I_2$ ) the lens defined by the intersection of $B(R,a)$ ( $B(R,b)$ ) with $B(R,b)$ ( $B(R,a)$ ) (See Figure \ref{Int2}). Since $\Lambda(I_1\cap I_2)=0$,  
\begin{equation*}
\begin{split}
\Lambda( & B(R,a) \cap B(R,b)) 
=\Lambda(I_1)+\Lambda(I_2).
\end{split}
\end{equation*}
To compute $\Lambda(I_1)$, consider the transformation that rotates $\delta$-degrees the vector $(0,0,c)-(d_E(a,b),0,\tilde{c}))$, about the $x-z$ axis (see Figure \ref{Int2}), 
\begin{equation*}
R_\delta=\left( \begin{array}{ccc}
cos(\delta) & 0 & -sin(\delta) \\
0           & 1 & 0           \\
sin(\delta) & 0 & cos(\delta)  \end{array} \right).
\end{equation*}

\begin{figure}[t!]
\center
\includegraphics[trim = 0mm 0mm 0mm 0mm, clip, width=200pt]{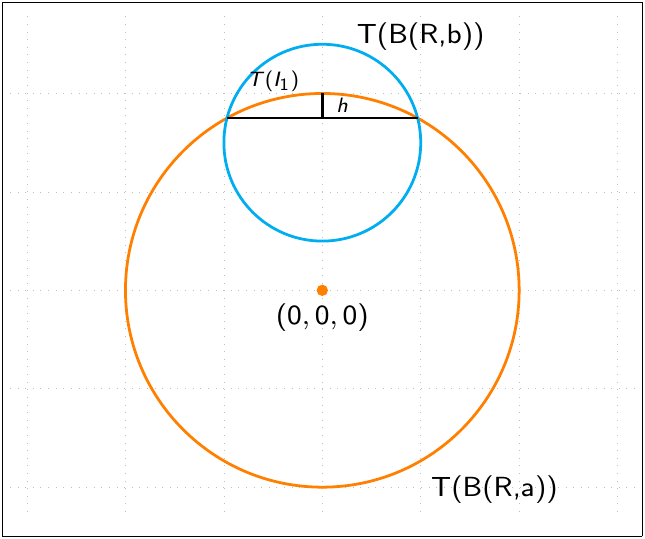}
\caption{The height $h$ of the lens $T(I_1)$, and the sets $T(B(R,a))$, and $T(B(R,a))$.}
\label{Int3}
\end{figure}

Given the strategic position of the centers, $\delta:= asin\left(\frac{c-\tilde{c}}{d}\right)+\frac{\pi}{2}$. 
Consider $\tau:\mathbb{R}^3 \rightarrow \mathbb{R}^3$, the translation sending $(0,0,c)$ to the origin, and define $T:\mathbb{R}^3 \stackrel{R_{\delta_1}}{\rightarrow} \mathbb{R}^3 \stackrel{\tau}{\rightarrow} \mathbb{R}^3 $. Given that $|DT^{(-1)}|=1$, the Lebesgue change of variable Theorem \cite[Th. 2.26]{RealComAnRud} states that
\begin{equation*}
\begin{split}
\Lambda(I_1) & = \iiint\limits_{I_1}\lambda f(w) dxdydw \\
& = \iiint\limits_{\tilde{T(I_1)}}\lambda f(c+wcos(\delta)-xsin(\delta)) dxdydw, \\
\end{split}
\end{equation*}
where the last equality follows from the fact that 
\begin{equation*}
(x,y,w) \stackrel{T^{(-1)}}{\longmapsto} (xcos(\delta)+wsin(\delta),y,c+wcos(\delta)-xsin(\delta))
\end{equation*} 
Remark that we can describe $T(I_1)$ in cylindrical coordinates (see Figure \ref{Int3}). Therefore,
\begin{equation*}
\begin{split}
& \Lambda(I_1)  \\
& =  \lambda \int\limits^{r}_{r-h} \int\limits^{2\pi}_0 \int\limits^{\sqrt{r^2-w^2}
}_0 \text{ \footnotesize $f(c+wcos(\delta)-scos(\theta)sin(\delta))$ } s ds d\theta dw,
\end{split}
\end{equation*}
where $h=\frac{(\tilde{r}-r+d)(r+\tilde{r}-d)}{2d}$ is the height of the lens defined by $I_1$ (see Appendix A). In the same fashion, 
\begin{equation*}
\begin{split}
& \Lambda(I_2)  \\
& =  \lambda \int\limits^{\tilde{r}}_{\tilde{r}-\tilde{h}} \int\limits^{2\pi}_0 \int\limits^{\sqrt{\tilde{r}^2-w^2}
}_0 \text{ \footnotesize $f(\tilde{c}+wcos(\tilde{\delta})-scos(\theta)sin(\tilde{\delta}))$ } s ds d \theta dw.
\end{split}
\end{equation*}

\subsection{Proof of Theorem \ref{ExpSin}}
Since the nearest neighbor is unique, for every atom $a\in \Phi$, we have that $\textbf{1}_{ \{ a\in \Phi^{(1)}\} }=\sum_{ b\in \Phi\backslash \{a\} }  \left( 1-\textbf{1}_{\{a \stackrel{\Phi,\mathcal{D}}{\leftrightarrow } b\}} \right)$, $\mathbb{P}$-a.s. Then, 
\begin{equation*}
\begin{split}
\mathcal{I}^{(1)}_g & := \sum_{a \in \Phi^{(1)}} g(\hat{a}), \\
& = \sum_{a \in \Phi}  \sum_{ b\in \Phi\backslash \{a\} }  g(\hat{a}) \left( 1-\textbf{1}_{\{a \stackrel{\Phi,\mathcal{D}}{\leftrightarrow } b\}} \right) , \ \ \mathbb{P}-a.s. \\
\end{split}
\end{equation*}
Using Campbell-Little-Mecke formula and Slivnyak-Mecke Theorem \cite{BacBlaVol1}, 
\begin{equation*}
\begin{split}
\mathbb{E} \left[ \mathcal{I}^{(1)}_g \right] & =  \int\limits_{\mathbb{H}^3}\int\limits_{\mathbb{H}^3} \mathbb{E}^{a,b} \left[ g(\hat{a}) \left(1-\textbf{1}_{\left\{a \stackrel{\Phi,\mathcal{D}}{\leftrightarrow} b \right\}} \right)  \right] \Lambda(db)\Lambda(da) \\
 & =  \int\limits_{\mathbb{H}^3} g(\hat{a}) \int\limits_{\mathbb{H}^3} \left(1- \mathbb{P}^{a,b} \left( a \stackrel{\Phi,\mathcal{D}}{\leftrightarrow} b \right) \right)  \Lambda(db)\Lambda(da). \\
\end{split}
\end{equation*}
For two different atoms $a$ and $b$, with marks $z$ and $\tilde{z}$, respectively,
\begin{equation*}
1-\mathbb{P}^{a,b} \left( a \stackrel{\Phi}{\leftrightarrow} b \right)=1-e^{-\lambda F(d_E(a,b),z,\tilde{z})} \textbf{1}^{(z,\tilde{z})}_\mathcal{D},
\end{equation*}
where $F(s,z,\tilde{z})$ is defined in Theorem \ref{theo1}. After the change of variable to polar coordinates, 
\begin{equation*}
\begin{split}
\int\limits_{\mathbb{H}^3} & \left(1- \mathbb{P}^{a,b} \left( a \stackrel{\Phi}{\leftrightarrow} b \right) \right)  \Lambda(db) \\
& = 2\pi \lambda \int\limits^\infty_0 \int\limits^\infty_0  \left( 1-e^{-\lambda F(s,z,\tilde{z})} \textbf{1}^{(z,\tilde{z})}_\mathcal{D} \right) sds f(\tilde{z}) d\tilde{z}  \\
& = 2\pi \lambda \int\limits^\infty_0 \left( 1-\mathbb{E} \left[e^{-\lambda F(s,z,\tilde{Z})} \textbf{1}^{(z,\tilde{Z})}_\mathcal{D}\right] \right) s  ds,
\end{split}
\end{equation*}
where the last equality follows after considering a change or order of integration, and a random variable $\tilde{Z}$, with distribution $f(z)dz$. In the same fashion, consider a random variable $Z$, independent of $\tilde{Z}$, with distribution $f(z)dz$, then 
\begin{equation*}
\begin{split}
& \mathbb{E} \left[ \mathcal{I}^{(1)}_g \right]  \\
& = 2\pi \lambda^2  \int\limits_{\mathbb{R}^2}  g(\hat{a}) 
\int\limits^\infty_0 \left( 1-\mathbb{E} \left[e^{-\lambda F(s,Z,\tilde{Z})} \textbf{1}^{(Z,\tilde{Z})}_\mathcal{D}\right] \right) s ds
   d\hat{a} \\
& = 2\pi \lambda^2 \int\limits_{\mathbb{R}^2} g(\hat{a})d\hat{a} \int\limits^\infty_0  \left( 1-\mathbb{E} \left[e^{-\lambda F(s,Z,\tilde{Z})} \textbf{1}^{(Z,\tilde{Z})}_\mathcal{D}\right] \right)s  ds.
\end{split}
\end{equation*}
Again, using Campbell-Little-Mecke formula and Slivnyak-Mecke Theorem \cite{BacBlaVol1}
\begin{equation*}
\begin{split}
\mathbb{E} \left[ \mathcal{I}^{(2)}_k \right] & = \frac{1}{2} \int\limits_{\mathbb{H}^3}\int\limits_{\mathbb{H}^3} \mathbb{E}^{a,b} \left[ k(\hat{a},\hat{b}) \textbf{1}_{\left\{a \stackrel{\Phi}{\leftrightarrow} b \right\}}  \right] \Lambda(da)\Lambda(db) \\
& = \frac{1}{2} \int\limits_{\mathbb{H}^3} \int\limits_{\mathbb{H}^3} k(\hat{a},\hat{b}) \mathbb{P}^{a,b} \left(a \stackrel{\Phi}{\leftrightarrow} b \right) \Lambda(da)\Lambda(db) \\
& = \frac{\lambda^2}{2} \int\limits_{\mathbb{R}^2} \int\limits_{\mathbb{R}^2}  k(\hat{a},\hat{b})\mathbb{E}\left[ e^{-\lambda F(d_E(\hat{a},\hat{b}),Z,\tilde{Z})} \textbf{1}^{(Z,\tilde{Z})}_\mathcal{D}\right]  d\hat{a} d\hat{b}.
\end{split}
\end{equation*}

\subsection{Proof of Theorem \ref{proporDoub}}
Consider $k(\hat{a},\hat{b}) =2 \textbf{1}_{\hat{a}\in A}$. Applying Theorem \ref{ExpSin} for this particular choice of $k(\hat{a},\hat{b})$, 
\begin{equation*}
\begin{split}
& \mathbb{E} \left[ \mathcal{I}^{(2)}_g \right] \\
 & = \lambda^2 \int\limits_{\mathbb{R}^2}\textbf{1}_{\hat{a}\in A} \left( \int\limits_{\mathbb{R}^2} \mathbb{E}\left[ e^{-\lambda F(d_E(\hat{a},\hat{b}),Z,\tilde{Z})} \textbf{1}^{(Z,\tilde{Z})}_\mathcal{D}\right] d\hat{b} \right) d\hat{a}.
\end{split}
\end{equation*}
After the change of variable to polar coordinates, 
\begin{equation*}
\begin{split}
\mathbb{E} & \left[ \mathcal{I}^{(1)}_h \right] \\
& =  \lambda^2  \int\limits_{\mathbb{R}^2} \textbf{1}_{\hat{a}\in A} \left( 2 \pi  \int\limits^\infty_0 \mathbb{E}\left[ e^{-\lambda F(s,Z,\tilde{Z})}\textbf{1}^{(Z,\tilde{Z})}_\mathcal{D} \right] s ds \right) d\hat{a} \\
& = \lambda \int\limits_{\mathbb{R}^2} \textbf{1}_{\hat{a}\in A} d\hat{a} \left( \lambda 2 \pi  \int\limits^\infty_0 \mathbb{E}\left[ e^{-\lambda F(s,Z,\tilde{Z})}\textbf{1}^{(Z,\tilde{Z})}_\mathcal{D}\right] s ds \right)\\
& =  \lambda \mathcal{S}(A) \left( \lambda 2 \pi  \int\limits^\infty_0 \mathbb{E} \left[ e^{-\lambda F(s,Z,\tilde{Z})}\textbf{1}^{(Z,\tilde{Z})}_\mathcal{D} \right] s ds\right).
\end{split}
\end{equation*}
Define 
\begin{equation*}
P_{\mathcal{D}}(\lambda,f):= \lambda 2 \pi  \int\limits^\infty_0 \mathbb{E}\left[ e^{-\lambda F(s,Z,\tilde{Z})}\textbf{1}^{(Z,\tilde
{Z})}_\mathcal{D} s ds\right].
\end{equation*}
Considering equation \eqref{M22}, we have that  
\begin{equation*}
M^{(2)}(A)= P_{\mathcal{D}}(\lambda,f) \lambda \mathcal{S}(A).
\end{equation*}
Since $$\lambda \mathcal{S}(A)=M^{(1)}(A)+M^{(2)}(A),$$
the following identity holds 
\begin{equation*}
M^{(1)}(dxdy)=(1- P_{\mathcal{D}}(\lambda,f)) \lambda dx dy.
\end{equation*}
We only have left to prove that $P(\lambda,f)$ lies in $[0,1]$. From the expression for $P(\lambda,f)$ , it is clear that it is always positive. Further, 
\begin{equation*}
\begin{split}
P_{\mathcal{D}}(\lambda,f) \lambda \mathcal{S}(A) = M^{(2)}(A) 
& \leq  \lambda \mathcal{S}(A),
\end{split} 
\end{equation*}
for every $A\in \mathcal{B}(\mathbb{R}^2)$. This implies that $P_{\mathcal{D}}(\lambda,f)\leq 1$.

\bibliographystyle{unsrt}
\footnotesize
\bibliography{FixClust}

\end{document}